\documentclass{webofc}

\usepackage[varg]{txfonts}   
\usepackage{hyperref}
\usepackage{url}
\hypersetup{colorlinks=true,citecolor=blue,urlcolor=blue,linkcolor=blue}
\usepackage{lineno}
\usepackage{xspace}
\newcommand{\s}            {\ensuremath{\sqrt{s}}\xspace}
\newcommand{\snn}          {\ensuremath{\sqrt{s_{\mathrm{NN}}}}\xspace}
\newcommand{\pt}           {\ensuremath{p_{\rm T}}\xspace}

\newcommand{\mee}          {\ensuremath{m_{\mathrm{ee}}}\xspace}

\newcommand{\gev}          {\ensuremath{\mathrm{GeV}}\xspace}
\newcommand{\gevc}          {\ensuremath{\mathrm{GeV}/c}\xspace}
\newcommand{\GeVc}          {\gevc}

\newcommand{\gevcc}         {\ensuremath{\mathrm{GeV}/c^{2}}\xspace}

\newcommand{\vtwo}          {\ensuremath{v_{2}}\xspace}
\newcommand{\mub}           {\ensuremath{\mu_{\mathrm{B}}}\xspace}
\newcommand{\dNchdeta}      {\ensuremath{\mathrm{d}N_{\mathrm{ch}}/\mathrm{d}\eta}\xspace}
\begin{document}
\title{Experimental overview of electromagnetic radiation in heavy-ion collisions\thanks{Invited plenary talk at the 31st International Conference on Ultra-relativistic Nucleus-Nucleus Collisions (Quark Matter 2025), Frankfurt, Germany, Apr 6-12, 2025.}}
%
%

\author{\firstname{H. Sebastian} \lastname{Scheid}\inst{1}\fnsep\thanks{\email{s.scheid@cern.ch}} }

\institute{CERN, Geneva, Switzerland}

\abstract{
Electromagnetic (EM) probes such as photons and dileptons provide direct insight into the space-time evolution of the hot and dense matter formed in heavy-ion collisions. Being unaffected by strong interactions, they serve as penetrating messengers from all collision stages, from pre-equilibrium dynamics to the quark–gluon plasma (QGP) and hadronic phases. This contribution summarises recent experimental results on direct-photon and dilepton production from RHIC and LHC experiments, as well as at lower energies. Particular emphasis is given to the ongoing “direct-photon puzzle,” the study of universal scaling of direct-photon production over a large range of collision systems and energies. Recent dielectron measurements from ALICE, STAR, and HADES, as well as new experimental developments at the LHC, are presented, along with perspectives for future facilities.
}
\maketitle
\section{Introduction}
\label{intro}
Electromagnetic radiation offers a unique probe into the dynamics of heavy-ion collisions, as photons and dileptons escape the QCD medium largely undisturbed. As a consequence, their production reflects properties of stages of the collision prior to hadronisation. In the earliest phase, the hard initial collisions produce prompt photons and while the system approaches equilibrium it radiates photons and dilepton. The measurement of thermal radiation in the QGP phase can be used to infer the average temperature of the medium prior to the transition into a hadronic state of matter. The latter also radiates thermally and can be used to study the in-medium EM spectral function and chiral-symmetry restoration.
In the following, I will present the most recent results on the measurements of real and virtual direct photons, i.e. photons not originating from hadron decays. An effort is made to relate the results to each other as well as to discuss them in the context of previous measurements. The discussion starts with real direct photons, followed by measurements of dileptons. The contribution concludes with remarks on the interpretation of the results in light of recent theoretical developments, which are discussed in detail in~\cite{Jackson2025}.

\section{Real direct photons}
\label{sec-1}
\begin{figure}[ht]
    \centering
    \sidecaption
    \includegraphics[width=0.40\textwidth]{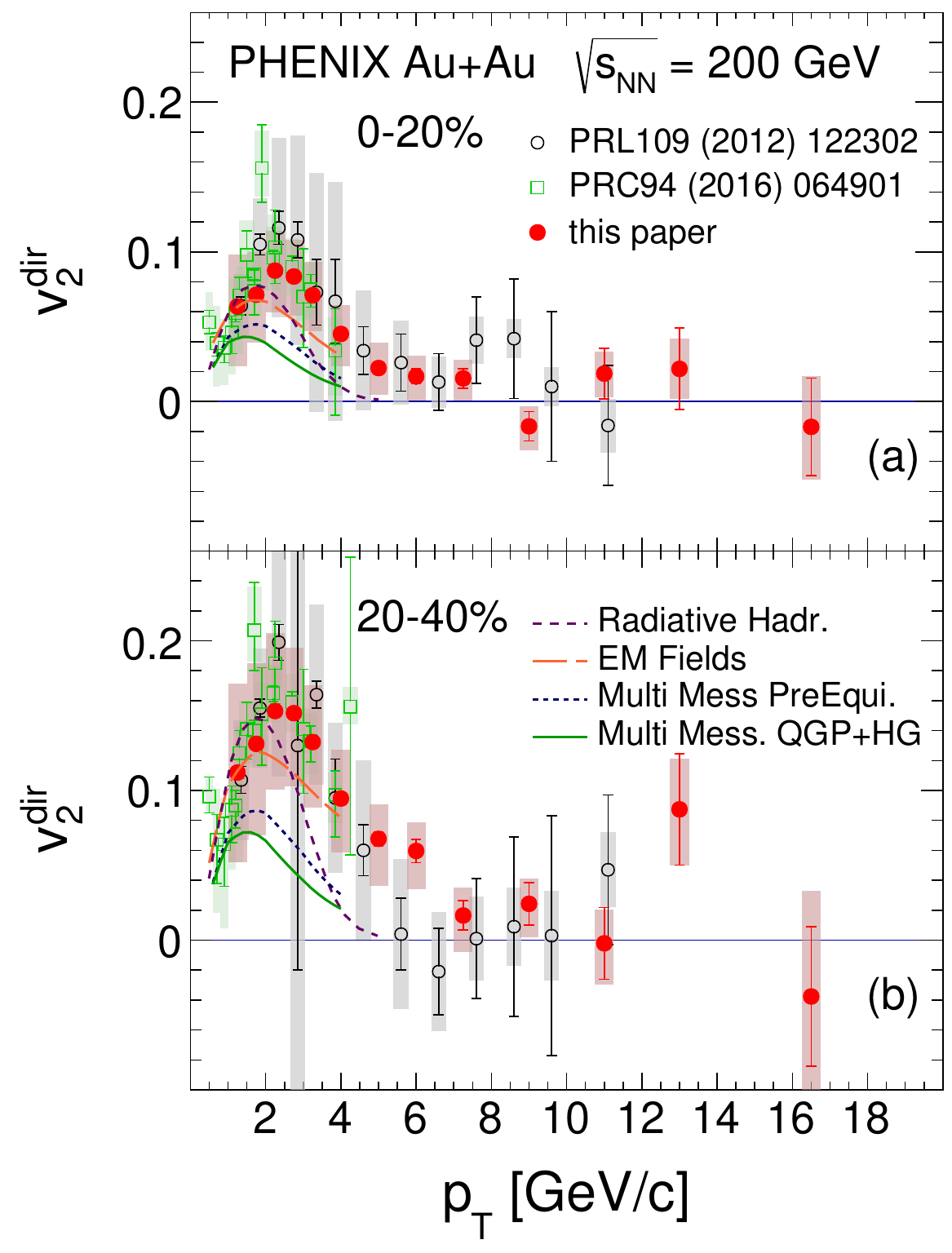}
    \caption{Measurement of the direct-photon elliptic flow \vtwo as a function of \pt by the PHENIX collaboration~\cite{PHENIX:2025ejr} in Au+Au collisions at \snn = 200 \gev. The measurement is presented in the centrality intervals of 0-20\% and 20-40\% and is compared to results from previous analyses in both cases the \pt reach is extended significantly. The data are compared to different model calculations.}
    \label{fig:phenixv2}
\end{figure}

Direct-photon production is a key observable in heavy-ion physics, yet the field faces the long-standing 'direct-photon puzzle'. The puzzle is rooted in the question of when the majority of the photons observed are produced. The understanding is that most photons must be produced when the medium is hot, so in an early phase of the collision. Photon production early in the collision would also imply a small elliptic flow (\vtwo) as the collectivity did not have the time to build up yet. Measurements of both observables by PHENIX in Au+Au collisions at $\snn = 200$~GeV posed a challenge to models in the simultaneous description of the yield and \vtwo.
New results from PHENIX based on the analysis of Au+Au data confirm the previous observation~\cite{PHENIX:2025ejr}. The large data sets taken in 2014 made it possible to reduce the uncertainties in the data and extend the range \pt of the \vtwo measurements as shown in Fig.~\ref{fig:phenixv2}, showing that at high \pt no elliptic flow is observed as expected from the explanation above.
The new data are also compared to a holistic state-of-the-art model that takes into account the different collision stages discussed before~\cite{Gale:2021emg}, however, the picture remains: the yield and \vtwo cannot be described in the same calculation. A way out is brought up by the PHENIX collaboration, which suggests a shift in the paradigm: large yield means early production and large \vtwo is rooted in a late production. The suggestion is that exotic sources like early electromagnetic fields could induce \vtwo in the early phase, while radiative processes in the hadronisation would produce additional photons late in the collision. As shown in figure~\ref{fig:phenixv2} both approaches give an enhancement of the \vtwo that would mitigate the tension that is observed between other models and the data.
It should be stressed that, while the sources labelled are exotic, are presumably reasonable contributions to flow and direct photons, it is highly non-trivial to constrain them. Factors such as the lifetime and strength of the EM fields in the early collision phases remain unknown.

\begin{figure}[ht!]
    \sidecaption
    \includegraphics[width=0.6\textwidth]{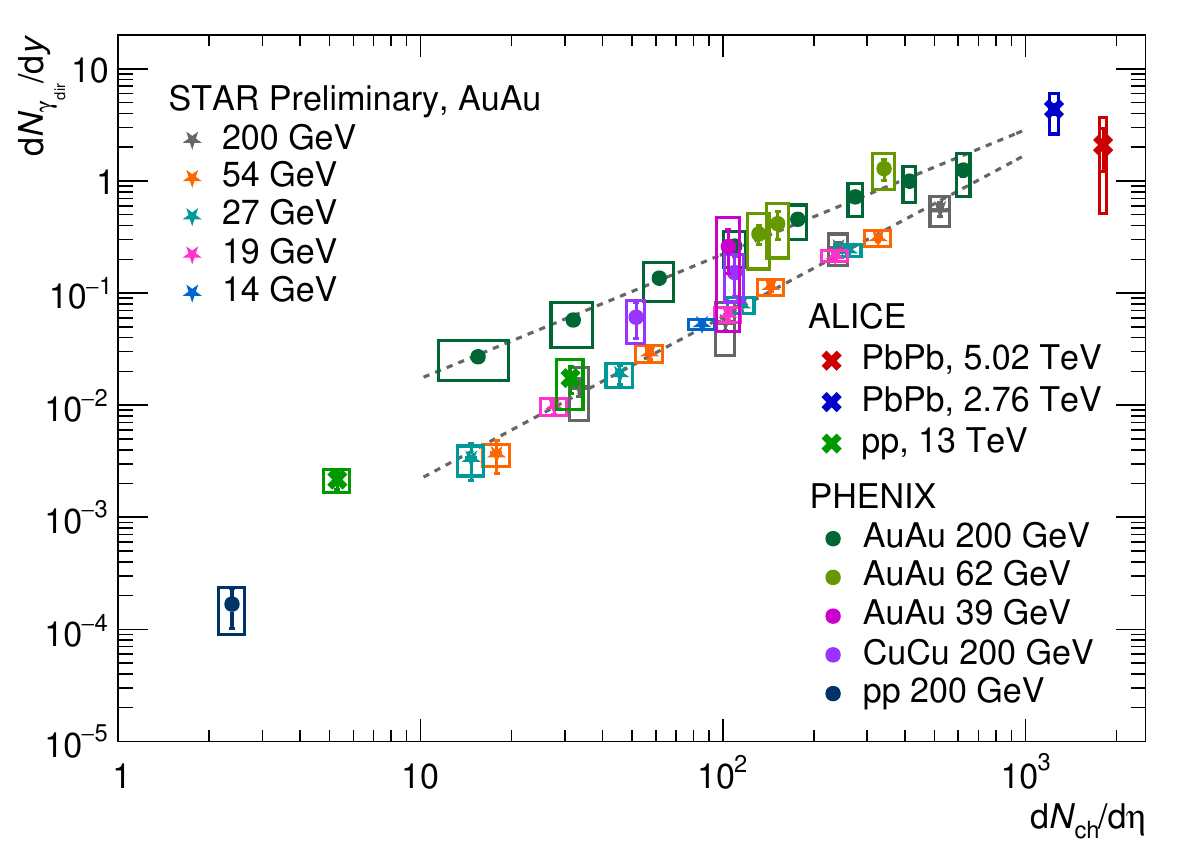}
    \caption{Direct-photon yield measured in different collision systems, energies and centralities as a function of the charged particle density \dNchdeta. The PHENIX~\cite{PHENIX:2022rsx} and ALICE~\cite{ALICE:2024vwy} yields are extracted in $1 < \pt < 5$ \GeVc, STAR~\cite{Bao2025} used $1 < \pt < 3$ \GeVc. The dashed lines indicate the universal scaling parameters extracted based on the latest PHENIX data ($\alpha \approx 1.1$) and the preliminary STAR data ($\alpha \approx 1.43$).}
    \label{fig:scaling}
\end{figure}

A new piece to the puzzle is the so-called universal scaling of the direct-photon production. The PHENIX collaboration first observed that the low-\pt direct-photon yield as a function of the charged particle density (\dNchdeta) can be described by a function $A \times (\dNchdeta)^{\alpha}$ and a universal $\alpha$ is found that describes the production independent of collision energies and centrality.
A new result was presented by the PHENIX collaboration~\cite{PHENIX:2022rsx} based on the same Au+Au dataset mentioned above which finds a value of $\alpha \approx 1.1$. A similar analysis was carried out by the STAR collaboration~\cite{Bao2025} which confirms a scaling of the direct-photon production that seems independent on the collision energy or centrality and only follows \dNchdeta. However, the measurement by STAR suggests a different power in the scaling law ($\alpha \approx 1.43$). In Fig.~\ref{fig:scaling} a summary of the STAR and PHENIX data is complemented by measurements of ALICE.

The dashed lines indicate the derived scaling with different $\alpha$ parameters extracted from fitting the STAR and the PHENIX data. The present state of the data is puzzling in two different ways. On the one hand, the obvious difference between the measurements of STAR and PHENIX poses an experimental challenge that needs to be resolved in the future. While the data from LHC provided by ALICE is not conclusive with the current uncertainties, new measurements with higher precision could help.
However, both STAR and PHENIX observe the scaling, and it is not clear why the photon yield should exhibit a universal scaling with \dNchdeta in the first place. In a simple picture, one can derive different scaling factors for the direct-photon yield depending on the collision phase ($\alpha_{\rm QGP} \approx 1.85, \alpha_{\rm hadronic}\approx 1.25$~\cite{Shen:2013vja}). In general, it could be expected that the change of collision energy and centrality changes the relative contributions of the photons from different phases of the collision, which makes it surprising that the yield can be described by this simple scaling. A possible interpretation is that the bulk of photons is produced at the transition of the system from the partonic to the hadronic phase. Here, one can expect that the energy density is the same for all collision systems and energies, while the system's volume varies and is the only degree of freedom to change the overall photon yield. 

While apparently the production of direct photons seems to follow a universal scaling in AA collisions, a special interest is in the onset of this phenomenon. For this a detailed study of pp collisions can be used to test the onset of thermal production of photons. A measurement of the fraction of direct photons ($r$) as a function of \pt in minimum-bias and high multiplicity pp collisions at $\s = 13$~TeV was performed by ALICE~\cite{ALICE:2024vwy}.
\begin{figure*}[ht!]
    \centering
    \begin{minipage}{0.47\textwidth} 
    \includegraphics[width=1.0\textwidth]{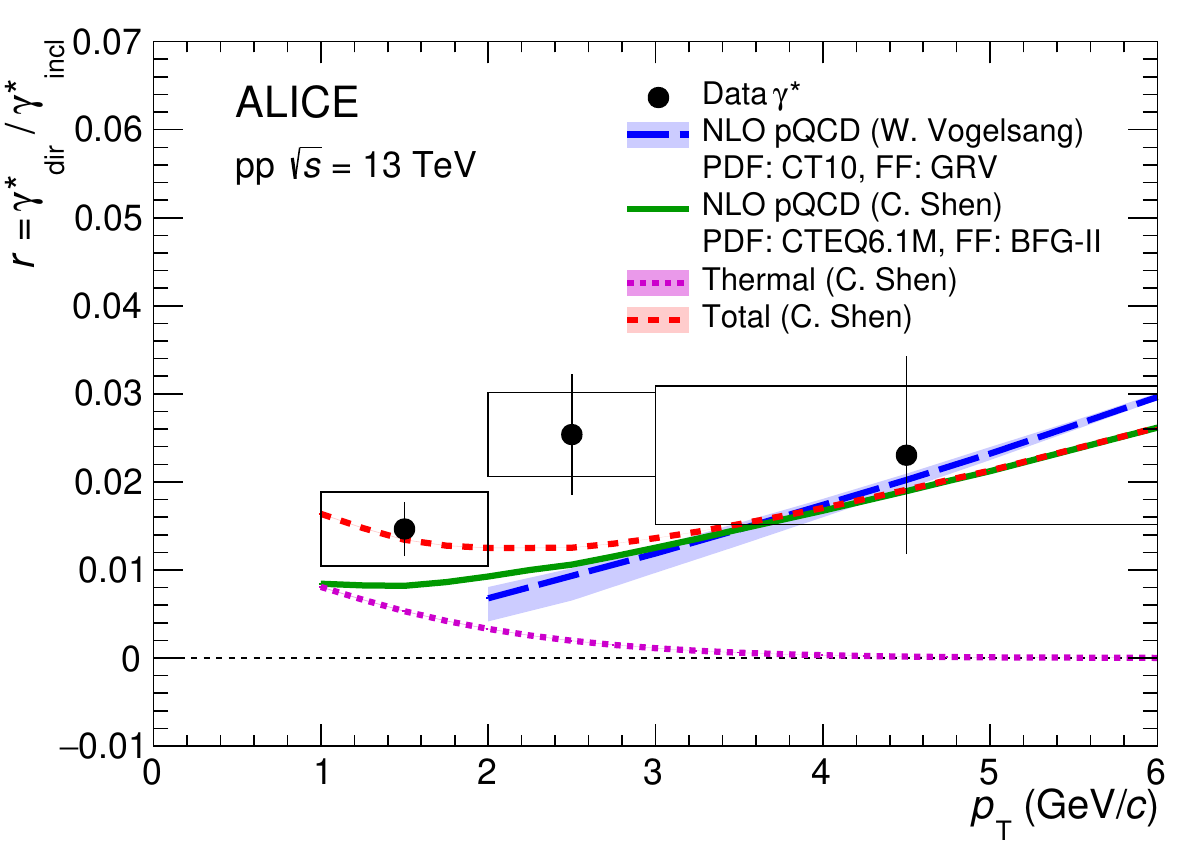}
    \end{minipage}
    \begin{minipage}{0.47\textwidth} 
    \includegraphics[width=1.0\textwidth]{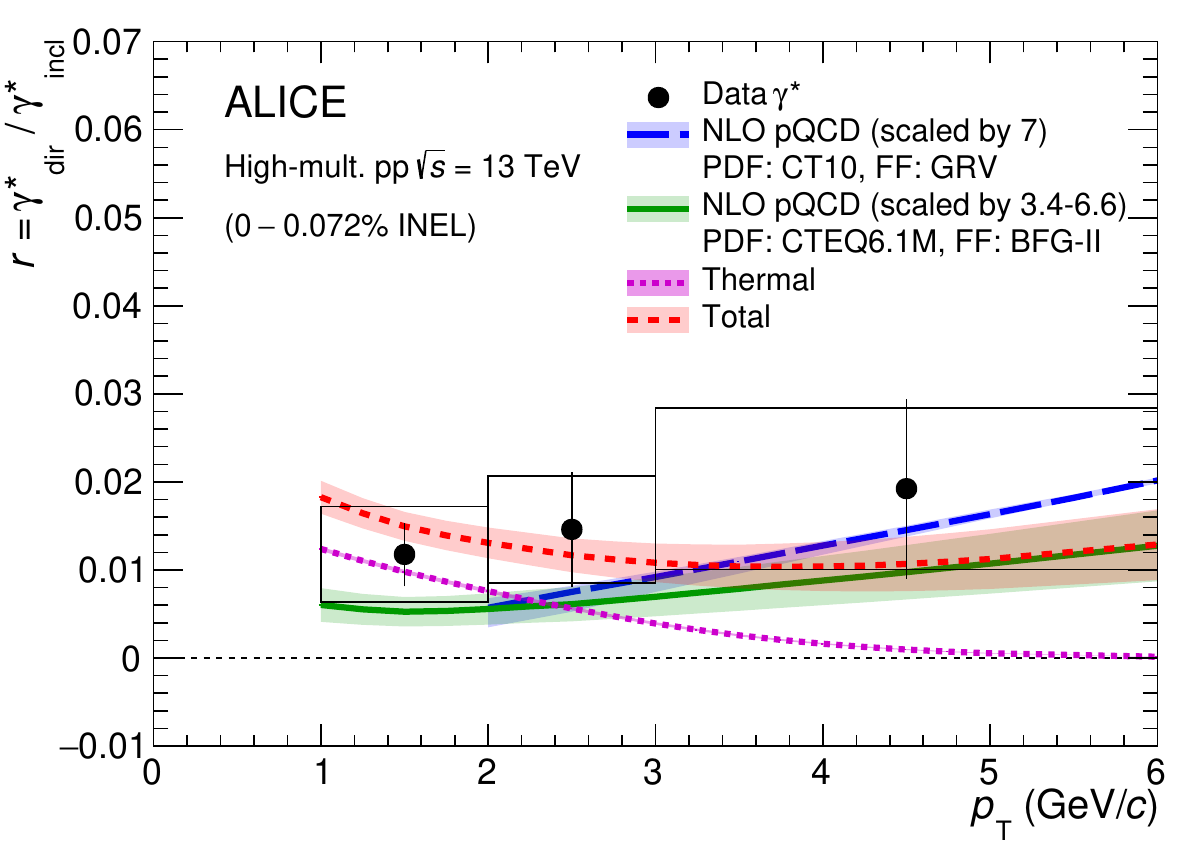}
    \end{minipage}
    \caption{The fraction of direct photons with respect to all photons ($r$) is presented as a function of $p_{\rm T}$ for minimum bias (left) and high-multiplicity (right) pp collisions at $\sqrt{s} = 13$~TeV measured with ALICE~\cite{ALICE:2024vwy}.}
    \label{fig:r-pp}
\end{figure*}
Figure~\ref{fig:r-pp} (left) shows the result for the measurement in minimum-bias pp collisions which indicated that about 1-2\% of all measured photons are direct photons. The production of direct photons can be understood in means of NLO pQCD calculations as shown in the comparison with two such calculations, as both can describe the data. The same measurement was carried out in high-multiplicity pp collisions and is shown in Figure~\ref{fig:r-pp} (right). The multiplicity corresponds to approximately 5 times the minimum-bias multiplicity. What is surprising is the fact that $r$ is about 1-2\% as in the minimum-bias case. However, an interpretation of the data in means of model comparison is non-trivial, as no calculation for NLO pQCD photon production in high-multiplicity pp collisions is available. The best description of the data by a parameterisation using the functional shapes of thermal and prompt pQCD photons from theory suggests that a small amount of the photons could be originating from thermal production. However, the data is also compatible with a scaled pQCD calculation alone.

\section{Dileptons}
\label{sec-2}
\begin{figure*}[ht]
    \centering
    \begin{minipage}{0.42\textwidth} 
    \includegraphics[width=1.0\textwidth]{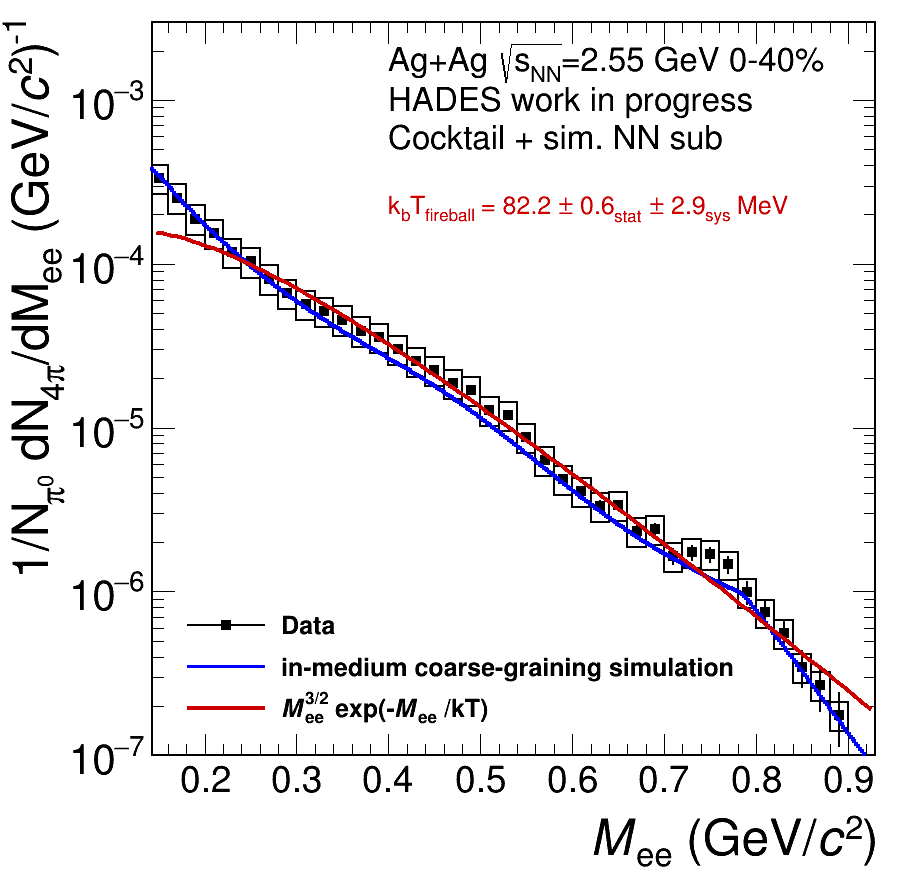}
    \end{minipage}
    \begin{minipage}{0.55\textwidth} 
    \includegraphics[width=1.0\textwidth]{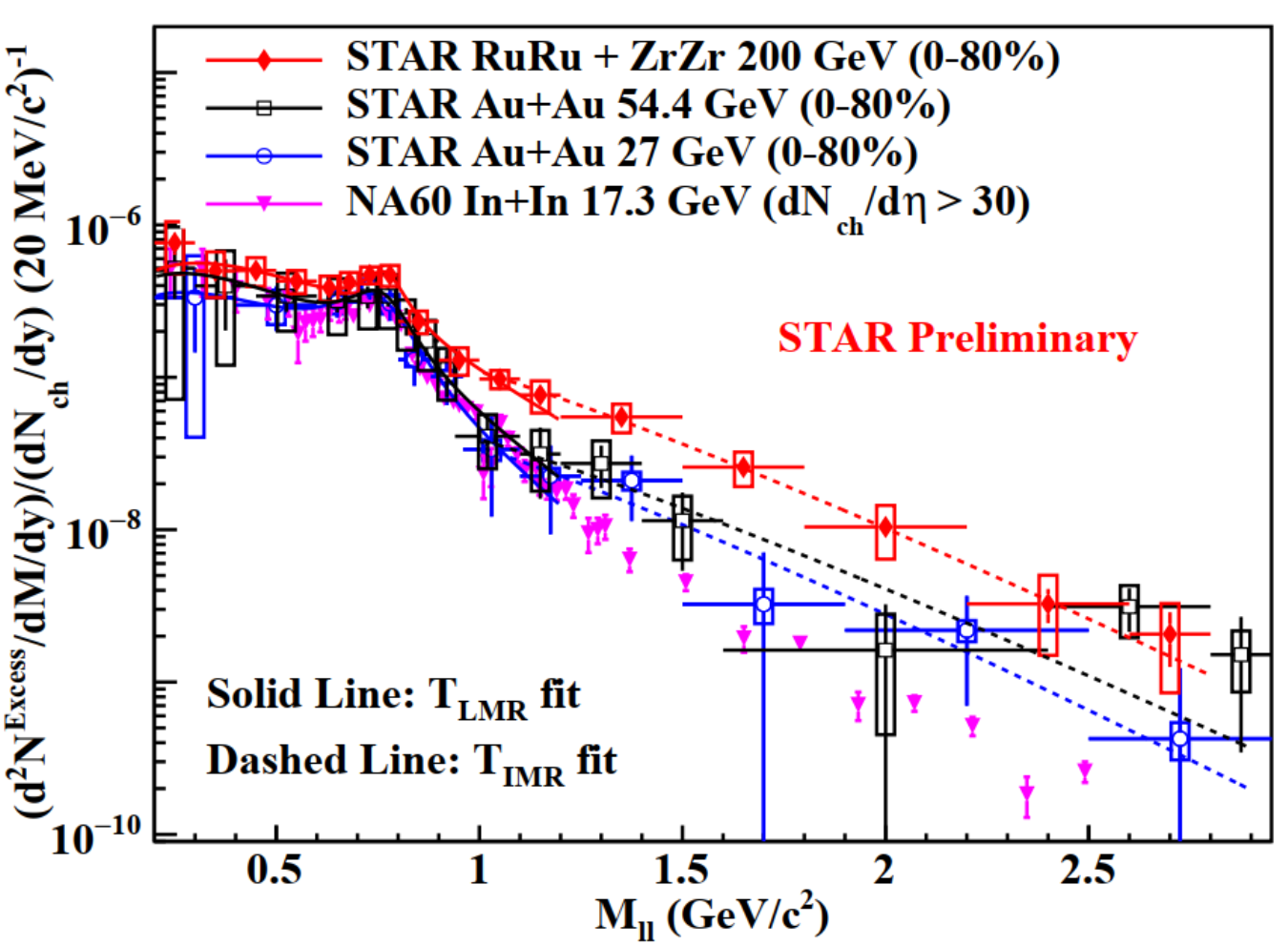}
    \end{minipage}
    \caption{Excess spectrum (after subtraction of hadronic contributions) measured by the HADES collaboration~\cite{HADES:2019auv,Schild2025} (left) and the STAR~\cite{STAR:2024bpc,Bao2025} and NA60~\cite{NA60:2008dcb} collaborations (right) in different collision systems and at varying centre-of-mass energies. The HADES data are compared to a in-medium spectral function based on coarse-grained simulations. Temperatures are extracted by parameterising the spectra as described in the text.}
    \label{fig:Texctraction}
\end{figure*}

While the measurements discussed before also use dielectrons to measure photon yields, the main advantage of dileptons is  a Lorentz invariant mass and thus, measurements related to the mass spectra are not subject to blue shifts and can give direct access to the temperature of the medium that radiates them.
The HADES collaboration presented preliminary results in Ag+Ag collisions at $\snn = 2.4$ and 2.55~GeV. The excess spectrum as a function of the dielectron invariant mass (\mee) is presented in Fig.~\ref{fig:Texctraction} (left)~\cite{Schild2025}.

\begin{figure}[ht]
    \sidecaption
    \includegraphics[width=0.5\textwidth]{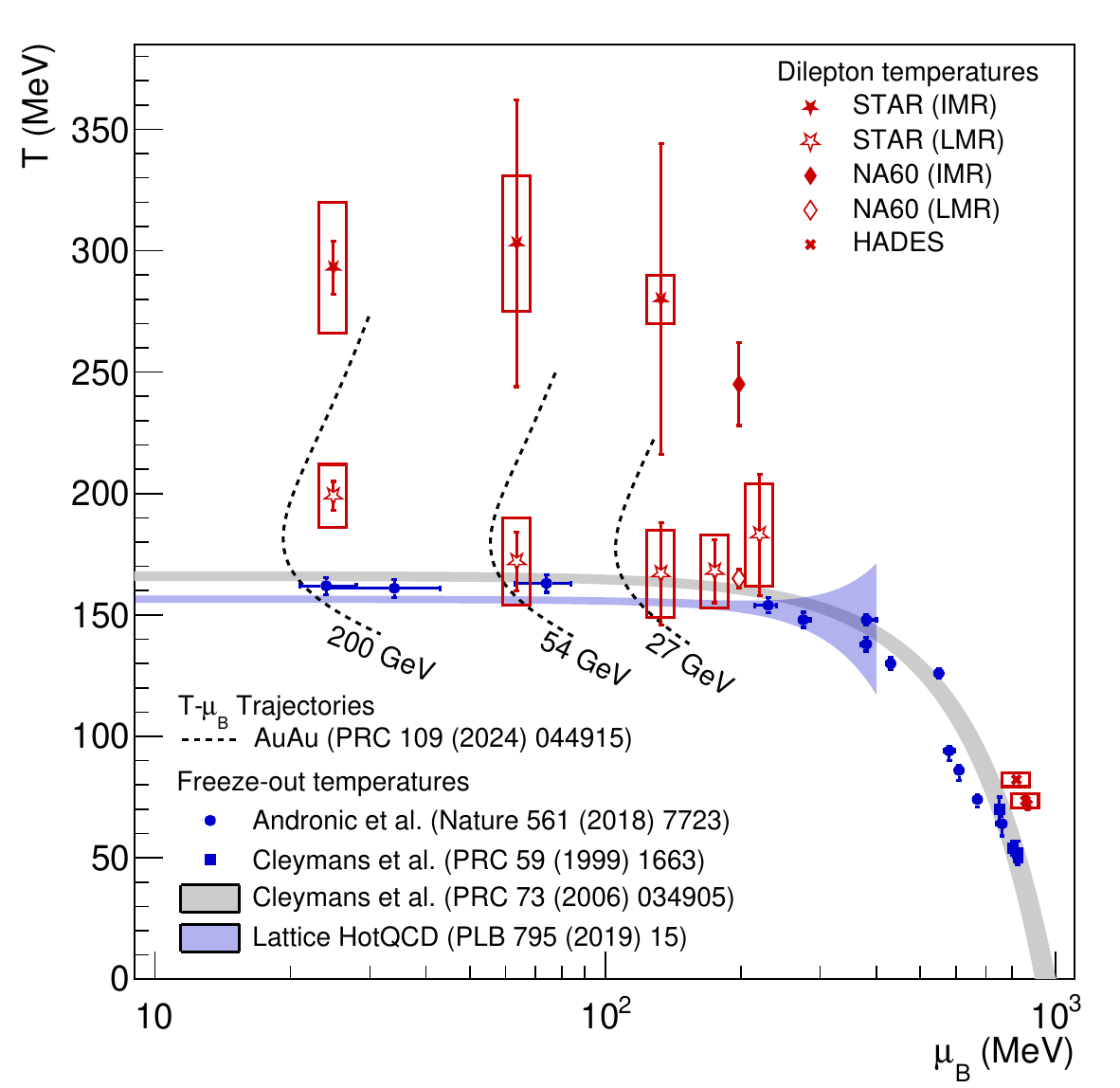}
    \caption{Temperatures extracted from the dilepton excess spectrum in the low-mass region (LMR, $0.4 < m_{\rm ll} < 1.2~\gevcc$) and the intermediate-mass region (IMR, $1 < m_{\rm ll} < 2.9~\gevcc$) based on measurements by the STAR~\cite{STAR:2024bpc,Bao2025} and NA60~\cite{NA60:2008dcb} collaborations as we all by HADES~\cite{HADES:2019auv,Schild2025} are presented in the $T$ vs $\mub$ plane. The data are shown together with freeze-out temperatures extracted in thermal model fits as well as predictions of the freeze-out temperature as a function of \mub. Trajectories based on hydrodynamic calculations of Au+Au collisions at different energies indicate  the path of a cooling system in the $T$ vs \mub plane. For more details see text.}
    \label{fig:phasediagram}
\end{figure}

The data is well described by an in-medium coarse-grained UrQMD calculation~\cite{Galatyuk:2015pkq}. The average fireball temperature is then extracted using the relation ${\rm d}N_{\rm ee}/{\rm d}\mee \propto \mee^{3/2} \exp(-\mee/T)$ and reported to be $82.2\pm0.6 {\rm(stat)} \pm 2.9{\rm(syst)}$~MeV. The STAR collaboration previously measured the dielectron access as a function of \mee in Au+Au collisions at \snn = 54.4 and 27~GeV~\cite{STAR:2024bpc}, in Fig.~\ref{fig:Texctraction} (right) the Au+Au data is presented together with a new preliminary measurement in Ru+Ru and Zr+Zr collisions at 200~GeV~\cite{Bao2025} as well as the measurement by the NA60 collaboration in In+In collisions at 17.3~GeV via dimuons. The excess is measured in all systems in the mass range from 0.2 to 2.9~\gevcc in Fig.~\ref{fig:Texctraction} (right). In contrast to the HADES data at low masses (<1~\gevcc) the structure of the spectral function is not completely gone in the STAR data and the remnants of the $\rho$ peak are present. To better compare the excess spectra in the different collision systems, the spectra are normalised by the charged-particle multiplicity at mid rapidity.
The STAR collaboration extracted temperatures in two distinct mass regions, $0.4 < m_{\rm ll} < 1.2$~\gevcc and $1 < m_{\rm ll} < 2.9$~\gevcc. While the parameterisation in the intermediate-mass region (IMR) is the same as used by the HADES collaboration, the lower mass region (LMR) includes an additional term of a relativistic Breit-Wigner function to accommodate the residual resonance structure. The temperatures extracted in all systems are summarized and compared to measurements and theoretical estimates of freeze-out temperature in Fig.~\ref{fig:phasediagram} as a function of the baryo-chemical potential \mub.

The different collision energies that cover 2.4 < \snn < 200~GeV give the possibility to probe the QCD phase diagram over a large range of \mub. 
It should be mentioned that while HADES used a theoretical model to extract \mub at the measured average temperature of the produced medium, the STAR data are plotted at the \mub that is extracted from thermal model fits in the same system and energy and corresponds to the \mub at freeze-out. To indicate the trajectory of the system in the $T-\mub$ plane calculations in Au+Au collisions at different energies are indicated as dashed lines.
While temperatures extracted in the LMR are close to the freeze-out temperature, indicating a system that is close to hadronic freeze-out the temperatures measured in the IMR are systematically above the critical temperature and can be interpreted as an average temperature of the QGP. A surprise in these measurements are the overall high temperatures. Comparing them with the shown trajectories, one would expect the average temperature to fall between the initial T of the system and the freeze-out temperature. The theoretical predictions do not include a possible pre-equilibrium radiation which could lead to an overall higher average temperature measurement.
Measurements of this sort are in particular sensitive to the subtraction of hadronic background stemming from semi-leptonic decays of charmed hadrons as well as the subtraction of Drell-Yan (DY) production of dileptons. While at the collision energies of several GeV the background from correlated HF decays are expected to be small, at LHC energies, they dominate the dilepton measurements over a wide mass range. The ALICE collaboration presented studies in which the separation of a prompt (resonance decays) and non-prompt (heavy-flavour decays) can be distinguished using a topological separation~\cite{Eisenhut2025}. 

These ALICE studies lay the ground to extend temperature measurements to LHC energies which in the phase diagram of QCD matter is pushing \mub to negligible values. Also measurements from LHCb in a different kinematic regime can be expected. With its capabilities of vertexing and muon identification, the experiment is well suited to study DY production and pre-equilibrium radiation at the LHC. In the further future the NA60+ experiment at the SPS at CERN and the CBM experiment at FAIR will bridge the phase space between measurements at RHIC and the results from the HADES collaboration to study dense nuclear matter under extreme conditions and explore the nature of the phase transition in this regime of the phase diagram. The next generation experiment at the LHC, ALICE 3, aims for precision studies of thermal dielectron production, such as the space time evolution of the QGP via differential temperature measurements and collectivity studies of dielectrons. Of particular interest of future dilepton experiments is the detailed line-shape of the in-medium spectral function and its connection to the partial restoration of chiral symmetry.

\section{Conclusions}
In the field of electromagnetic probes several new measurements were presented that will help in the fundamental understanding of the QGP produced in relativistic heavy-ion collisions. The photon puzzle is still present, and exotic approaches in theory may give possibilities for a resolution, however constraining these sources has proven challenging. A more pressing issue on the experimental side is the disagreement of data presented on the universal scaling of direct-photon production presented by PHENIX and STAR which makes the direct-photon puzzle only more puzzling. Data from LHC would be of great value for a possible solution of this discrepancy.
Dilepton measurements across a wide energy range allow extraction of fireball temperatures and mapping of QCD matter in the phase diagram. Results from the ALICE Run 3 program give an idea of possibilities to extract temperature measurements in the \mub = 0 regime.
The plans for future experiments at LHC, SPS and FAIR promise a bright future in the field of electromagnetic probes.

%

\begin{thebibliography}{}

\footnotesize
\bibitem{Jackson2025}
    G. Jackson,
    \href{https://inspirehep.net/literature/2965679}{\it Theory overview on electroweak emission from heavy-ion collisions}
    (these proceedings, 2025).

\bibitem{PHENIX:2025ejr}
    PHENIX,
    \href{https://inspirehep.net/literature/2907944}{\it Azimuthal anisotropy of direct photons in Au$+$Au collisions at $\sqrt{s_{_{NN}}}=200$ GeV}
    (arXiv, 2025).

\bibitem{Gale:2021emg}
    C. Gale et al.,
    \href{https://inspirehep.net/literature/2010837}{\it Multimessenger heavy-ion collision physics}
    (PRC 105, 014909, 2022).

\bibitem{PHENIX:2022rsx}
    PHENIX,
    \href{https://inspirehep.net/literature/2061074}{\it Nonprompt direct-photon production in Au+Au collisions at $\snn=200$ GeV}
    (PRC 109, 044912, 2024).

\bibitem{Bao2025}
    X. Bao,
    {\it Direct virtual photon measurements in Au+Au collisions with STAR BES-II data}
    (these proceedings, 2025).

\bibitem{ALICE:2024vwy}
    ALICE Collaboration,
    \href{https://inspirehep.net/literature/2850920}{\it Direct-photon production in inelastic and high-multiplicity proton{\textendash}proton collisions at $\sqrt{s}= 13$ TeV}
    (PLB 868, 139645, 2025).

\bibitem{Shen:2013vja}
    C. Shen et al.,
    \href{https://inspirehep.net/literature/1247475}{\it Thermal photons as a quark-gluon plasma thermometer reexamined}
    (PRC 89, 0444910, 2014).

\bibitem{Schild2025}
    N. Schild,
    {\it Studying properties of baryon-dominated matter with dileptons}
    (these proceedings, 2025).

\bibitem{HADES:2019auv}
    HADES,
    \href{https://inspirehep.net/literature/1758156}{\it Probing dense baryon-rich matter with virtual photons}
    (Nature Phys. 15, 1040, 2019).

\bibitem{STAR:2024bpc}
    STAR,
    \href{https://inspirehep.net/literature/2755369}{\it Temperature measurement of quark-gluon plasma at different stages},
    (arXiv, 2024).

\bibitem{NA60:2008dcb}
    NA60,
    \href{https://inspirehep.net/literature/799832}{\it Evidence for the production of thermal-like muon pairs with masses above 1 GeV/$c^{2}$ in 158 AGeV Indium-Indium Collisions},
    (EPJC 59, 607, 2009).
    
\bibitem{Galatyuk:2015pkq}
    T. Galatyuk et al.,
    \href{https://inspirehep.net/literature/1411749}{\it Thermal Dileptons from Coarse-Grained Transport as Fireball Probes at SIS Energies}
    (EPJA 52, 131, 2016).

\bibitem{Eisenhut2025}
    F. Eisenhut,
    \href{https://inspirehep.net/literature/2967351}{\it Dielectron production in pp and Pb-Pb collisions with ALICE in Run 3}
    (these proceedings, 2025).



















\end{thebibliography}

\end{document}